\documentclass[final,5p,times,twocolumn]{elsarticle}
\usepackage{eurosym}
\usepackage{graphicx}
\usepackage{amssymb,bm,amsmath}
\usepackage{indentfirst}
\usepackage{epsfig}
\usepackage{epstopdf}
\usepackage{exscale}
\usepackage{relsize}
\usepackage[colorlinks=true,citecolor=black]{hyperref}
\usepackage{caption2}

\journal{Chaos, Solitons \& Fractals}
\biboptions{sort&compress}
\biboptions{}

\begin{document}

\begin{frontmatter}

\title{Hopfions in the Lee-Huang-Yang superfluids}
%\author[a]{Liangwei Dong}

\author{Liangwei Dong \corref{cor1}}
\cortext[cor1]{Corresponding author.}
\ead{dlw0@163.com}
\address{Department of Physics, Zhejiang University of Science and Technology, Hangzhou,  310023, China}

\author{Mingjing Fan}
\address{The MOE Key Laboratory of Weak-Light Nonlinear Photonics, TEDA Applied Institute and School of Physics, Nankai University, Tianjin, 300457, China}

\author{Boris A. Malomed}
\address{Department of Physical Electronics, School of Electrical Engineering, Faculty of Engineering, Tel Aviv
University, Tel Aviv 69978, Israel}
\address{Instituto de Alta Investigacion, Universidad de Tarapaca, Casilla 7D, Arica, Chile}

\author{Yaroslav V. Kartashov}
\address{Institute of Spectroscopy, Russian Academy of Sciences, Troitsk, Moscow, 108840, Russia}
%
%
%\author{Boris A. Malomed}
%\address{Instituto de Alta Investigacion, Universidad de Tarapaca, Casilla
%7D, Arica, Chile\footnote{Sabbatical address}}
%\address[b]{aaDepartment of Physics, Shaanxi University of Science and Technology, Xi'an, China, 710021}
%
%

\date{\today}

%% To be edited by editor
% \dates{Compiled \today}

%\ociscodes{(140.3490) Lasers, distributed feedback; (060.2420) Fibers, polarization-maintaining;(060.3735) Fiber Bragg gratings.}

%% To be edited by editor
% \doi{\url{http://dx.doi.org/10.1364/XX.XX.XXXXXX}}

\begin{abstract}
	
 It is known that, under appropriate conditions, mean-field interactions can be canceled in binary BEC, leading to the formation of the Lee-Huang-Yang (LHY) superfluid, in which the nonlinearity is solely represented by the quartic LHY term. In this work we systematically investigate the existence, stability and evolution of hopfion states in this species of quantum matter. They are characterized by two independent topological winding numbers: inner twist $s$ of the vortex-ring core and overall vorticity $m$. The interplay between the LHY self-repulsion and a trapping harmonic-oscillator potential results in stability of the hopfions with $s = 1$ and $m$ ranging from $0$ to $4$. The hopfions exhibit distinct topological phase distributions along the vertical axis and the radial direction in the horizontal plane. Their effective radius and peak density increase with the chemical potential, along with expansion of the vortex-ring core. Although the instability domain of the hopfion modes broadens with the increase of $m$, stable hopfions persist in a wide range of the chemical potential, up to $m=4$, at least, provided that the norm exceeds a certain threshold value. The predictions are experimentally accessible in currently used BEC setups.

\end{abstract}


\end{frontmatter}

\section{Introduction}

\label{Sec1}

The creation of stable three-dimensional (3D) nonlinear states is a fundamental challenge across diverse areas of physics, including optics \cite{RevModPhys.83.247,malomed2022}, Bose-Einstein condensates (BECs) \cite{NaturePhys2009, ZHAO2024115329}, magnetics \cite{PhysRevLett.82.1554}, superconductors \cite{RevModPhys.95.025001}, and the classical field theory \cite{RADU2008101, PhysRevLett.131.111601}. The primary obstacle impeding direct creation of such states stems from the collapse instability \cite {PHYREP1998, Zakharov:2012, fibich2015singular}, driven by the ubiquitous cubic nonlinearity, manifested as the Kerr self-focusing in optics \cite{kivshar2003} and attractive contact interactions in ultracold atomic gases \cite{pitaevskii2003bose}.

In 1957, Lee, Huang, and Yang derived the leading-order beyond-mean-field (MF) correction for the single-component $3$D hard-sphere gas (i.e., with
repulsive interactions between particles) \cite{PhysRev.106.1135}. This correction, known as the Lee-Huang-Yang (LHY) term, quantifies the impact of
quantum fluctuations of the dynamics of the atomic BEC \cite {PhysRevLett.115.155302, PhysRevLett.117.100401}. In particular, the
correction plays a crucial role in stabilizing ultracold atomic gases in cases when beyond-MF effects become significant. Experimentally, LHY
physics, arising from weak quantum fluctuations, has been observed in various BEC atomic settings, including BECs trapped in optical lattices \cite
{PhysRevLett.81.3108, yang2020cooling}, quantum phase transitions from superfluid to Mott-insulator states \cite{greiner2002quantum}, and quantum
criticality \cite{zhang2012observation}.

Recently, the use of the LHY phenomenology has provided remarkable progress in theoretical and experimental studies of quantum droplets (QDs) \cite
{Nature2016_3, Science2018, PhysRevLett.120.135301, PhysRevLett.120.235301, Science2018_2,PhysRevA.98.013612,
Raymond-vortex,Raymond-discrete,Tylutki,DongPRL2021}. With the help of the Feshbach-resonance technique, the LHY and MF nonlinearities can be tuned to
comparable magnitudes but with opposite signs, enabling the formation of stable balanced states that avoid the collapse inherent in pure MF systems.
The interplay between contact repulsion and dipole-dipole attraction has led to the observation of anisotropic quantum droplets in dipolar gases \cite
{PhysRevLett.116.215301,NATU2016}. Subsequently, nearly 2D \cite{Science2018, PhysRevLett.120.135301} and 3D isotropic QDs \cite
{PhysRevResearch.2.013269} have been experimentally realized and theoretically investigated in mixtures of two atomic states in $^{39}$K.
Further, QDs have been created in attractive mixtures of $^{41}$K and $^{87}$Rb atoms with contact interactions \cite{PhysRevResearch.1.033155}, and in
binary dipolar BECs \cite{PhysRevLett.126.025301, PhysRevLett.126.025302}. Self-evaporation dynamics of quantum droplets was investigated in a $^{41}$K
and $^{87}$Rb mixture \cite{app11020866}. 

In addition to QDs stabilized by the interplay of the MF and LHY effects in binary BEC mixtures, J{\o }rgensen \emph{et al.} have predicted that, in the
binary condensates with atom numbers of the two components, $N_{1}$ and $N_{2}$, and the intra- and inter-component interaction coefficients, $%
g_{11,22}$ and $g_{12}$, precisely tuned to specific conditions, $\sqrt{g_{11}} N_{1}=\sqrt{g_{22}}N_{2}$ and $g_{12}=-\sqrt{g_{11}g_{22}}$, the
effective MF interactions entirely vanish due to mutual cancellation, leaving the system (LHY superfluid)\ governed solely by quantum fluctuations
\cite{PhysRevLett.121.173403}. An effective one-component Gross-Pitaevskii equation (GPE) accurately describes the dynamics of the LHY superfluids in this case, as confirmed by the comparison with the full two-component system. Experimentally, such 3D LHY superfluids have been realized in a $^{39}$K
spin mixture confined in a spherical trap \cite{PhysRevLett.126.230404}. Recently, the 1D counterpart of the LHY superfluid has also been studies
theoretically \cite{zengjianhua}. Nevertheless, so far only the simplest states with trivial phases were studied in LHY quantum superfluids. Here we
show that, in the presence of an external trapping potential, this state of quantum matter can as well support stable topological modes with a highly
nontrivial phase structure called hopfions.

Hopfions are 3D topologically organized states named after Heinz Hopf, who discovered the Hopf fibration in 1931 \cite{hopf1931abbildungen}. These 3D
states, which were originally identified in the field theory, exhibit robust particle-like characteristics. Inheriting the
topological properties of the Hopf fibration, they are defined by a homotopic mapping from a unit sphere in the 4D space to the 3D unit sphere.
Hopfions represent a special class of toroidal modes \cite{wan2022scalar}, which are characterized by two independent winding numbers: a hidden twist ($s$), which represents a circular vortex thread, and the overall vorticity around the vertical axis ($m$). Hopfions have been demonstrated in various
areas, including the field theory \cite{Faddeev,Manton}, magnetic materials \cite{magnet1,magnet2,zheng2023hopfion}, semiconductors/superconductors \cite{Babaev}, liquid crystals \cite{Smalyukh}, ferroelectrics \cite{luk2020hopfions, LUKYANCHUK20251}, and high-harmonic generation in optical
media \cite{lyu2024formation}.

Usually, hopfion states emerge in multicomponent systems or as quasi-linear modes confined by toroidal potentials. On the other hand, it was predicted
that hopfions can be stabilized in inhomogeneous media with the repulsive nonlinearity even without a linear potential, provided the nonlinearity
strength grows fast enough from the center to the periphery \cite{PhysRevLett.113.264101}. Stable toroidal modes with $s=1$ and vorticities $%
m=0,1$ have been predicted in the latter model.

While LHY superfluids were theoretically predicted \cite{PhysRevLett.121.173403} and experimentally observed \cite{PhysRevLett.126.230404}, their precise configuration and spatial structure remain poorly understood. The known stabilization mechanism for 3D vortex modes, with topological charges up to $m=8$, by means of the balanced LHY nonlinearity and trapping potential \cite{DONG2023113728} suggests two critical questions: (i) Can 3D twisted toroidal vortex states exist in LHY superfluids? (ii) What characterizes their dynamical evolution? Through numerical analysis of binary BECs trapped in a harmonic-oscillator (HO) trapping potential, we systematically investigate the existence, stability, and evolution of hopfions of this type. The findings demonstrate that the LHY-dominated hopfions undergo chemical-potential-dependent expansion, while maintaining their topological integrity. We show that hopfions with $s=1$ and $m$ values at least up to $4$ and sufficiently large number of particles exhibit exceptional stability, keeping the initial configurations without any observable deformation for an arbitrarily long time.

The following presentation is organized as follows. The model, based on the effective 3D GPE with the LHY term, is formulated in Section 2. Systematic
numerical results, which reveal stable hopfion families, are reported in Section 3. In the same section, we discuss conditions for the experimental
implementation of the predicted hopfion states. The paper is concluded by Section 4.

\section{The model}

\label{Sec2} In the MF approximation, a two-component BEC at zero temperature is modeled by the energy functional \cite{pethick2008bose}
\begin{equation}
E_{\text{MF}}=\mathlarger{\mathlarger{\int}}\left[ \mathlarger{\sum}_{i}\left( \frac{\hbar ^{2}|\nabla \Psi _{i}|^{2}}{2m_{i}}+R_{i}^{\prime
}(r)n_{i}\right) +\frac{1}{2}\mathlarger{\sum}_{ij}g_{ij}n_{i}n_{j}\right] \text{d}\mathbf{r},  \label{Eq1}
\end{equation}%
where subscripts $i=1,2$ denote the two components, $\Psi _{i}(\mathbf{r})=\sqrt{N_{i}}\psi _{i}(\mathbf{r})$ represents the wave function of
condensates with atom numbers $N_{i}$ and particle densities $n_{i}(\mathbf{r})\equiv |\Psi _{i}(\mathbf{r})|^{2}$, and $R_{i}^{\prime }(r)$ are trapping
potentials. The interatomic interactions are determined by the coupling constants $g_{ij}=2\pi \hbar ^{2}a_{ij}(m_{i}+m_{j})/(m_{i}m_{j})$, where $a_{ij}$ are the $s$-wave scattering lengths for collisions between atoms belonging to components $i$ and $j$. In the case of equal masses ($m_{1}=m_{2}\equiv m$) and identical HO trapping potentials [$R_{1}^{\prime }(\mathbf{r})=R_{2}^{\prime }(\mathbf{r})\equiv R^{\prime }(\mathbf{r})=m\omega _{0}^{2}r^{2}/2$], the MF energy functional simplifies to
\begin{equation}
\begin{aligned} E_{\text{MF}}&= \mathlarger{\mathlarger{\int}} \left[\frac{\hbar^2(|\nabla\Psi_1|^2+|\nabla\Psi_2|^2)}{2m}+R'(r)(n_1+n_2)
\right] \text{d}\textbf{r} \\ &+\mathlarger{\mathlarger{\int}} \left[\frac{4\pi\hbar^2}{m}(a_{11}n_1^2+a_{22}n_2^2+a_{12}n_1
n_2)\right]\text{d}\textbf{r}. \end{aligned}  \label{Eq2}
\end{equation}

The intra-species stability of the model based on Eq.~(\ref{Eq2}) is secured by positive intraspecies scattering lengths, i.e., $a_{11}>0$, $a_{22}>0$.
When the inter-species negative (attractive) scattering length reaches the critical value $a_{12}=-\sqrt{a_{11}a_{22}}$, one eigenvalue of the
quadratic form in Eq.~(\ref{Eq2}) is zero and the other is positive. The eigenvector corresponding to the zero eigenvalue has $n_{2}=\sqrt{a_{11}/a_{22}}n_{1}$. Therefore, the respective MF energy vanishes at this value of the density ratio, any deviation from it being energetically
costly. More specifically, when both conditions $a_{12}=-\sqrt{a_{11}a_{22}}$ and $N_{2}/N_{1}=\sqrt{a_{11}/a_{22}}$ are simultaneously satisfied, the MF
terms are completely canceled in Eq.~(\ref{Eq2}), and the corresponding strongly correlated condensate wavefunctions obey the constraint
\begin{equation}
\Psi _{2}=\Psi _{1}(a_{11}/a_{22})^{1/4}.  \label{PsiPsi}
\end{equation}%
In this case, the MF interaction vanishes, and any perturbation to one component leads to a restoring force towards relation (\ref{PsiPsi}).

If the relatively strong MF interaction is canceled, the residual dynamics is dominated by the weak repulsive LHY nonlinearity. The LHY contribution to
the energy density of the Bose mixture with equal masses $m$ is \cite{PhysRevLett.115.155302}
\begin{equation}
\frac{\varepsilon _{\text{LHY}}}{V}=\frac{32\sqrt{2\pi }\hbar ^{2}}{15m} \mathlarger{\sum}_{\pm }\left( a_{11}n_{1}+a_{22}n_{2}\pm \kappa \right)
^{5/2},  \label{Eq3}
\end{equation}%
with $\kappa \equiv \lbrack
(a_{11}n_{1}-a_{22}n_{2})^{2}+4a_{12}n_{1}n_{2}]^{1/2}$. Under the conditions $a_{12}=-\sqrt{a_{11}a_{22}}$ and $n_{1}/n_{2}=\sqrt{a_{22}/a_{11}%
}$, the energy density is reduced to
\begin{equation}
\frac{\varepsilon _{\text{LHY}}}{V}=\frac{256\sqrt{\pi }\hbar ^{2}}{15m}\left( n|a_{12}|\right) ^{5/2},  \label{Eq4}
\end{equation}%
where $n=n_{1}+n_{2}\equiv |\Psi _{1}(\mathbf{r})|^{2}+|\Psi _{2}(\mathbf{r})|^{2}$ is the total condensate density. Adding the above LHY correction to
Eq.~(\ref{Eq2}) and defining $|\Psi |^{2}\equiv |\Psi_{1}|^{2}+|\Psi_{2}|^{2}$, with $\Psi _{2}=\Psi _{1}(a_{11}/a_{22})^{1/4}$,
one obtains the effectively single-component energy functional
\begin{equation}
E=\mathlarger{\mathlarger{\int}}\left[ \frac{\hbar ^{2}|\nabla \Psi |^{2}}{2m}+R^{\prime }(r)\left\vert \Psi \right\vert ^{2}+\frac{256\sqrt{\pi }\hbar
^{2}}{15m}|a_{12}|^{5/2}|\Psi |^{5}\right] \text{d}\mathbf{r}.  \label{Eq5}
\end{equation}

The corresponding GPE, with the nonlinearity stemming solely from the LHY terms, is derived in the following form:
\begin{equation}
i\hbar \frac{\partial \Psi }{\partial t}=\left[ -\frac{\hbar ^{2}}{2m}\nabla^{2}+R^{\prime }(\mathbf{r})+\frac{128\sqrt{\pi }\hbar ^{2}}{3m}|a_{12}|^{5/2}|\Psi |^{3}\right] \Psi .  \label{Eq6}
\end{equation}%
Equation (\ref{Eq6}) is the model of the LHY\ superfluid in an interval around the value of the magnetic field which imposes the necessary Feshbach
resonance \cite{PhysRevLett.121.173403}. To cast the GPE in the scaled form,
we define characteristic units of length $r_{0}=\hbar/\sqrt{m \varepsilon _{0}}$, time $t_{0} = \hbar/\varepsilon_{0} = m r_{0}^{2}/\hbar$.  The characteristic factors for the normalization of the external potential and wave function are $R=R'/\varepsilon _{0}$ and $\psi=\Psi/\Psi_0$ with $\Psi_{0}= [ 3/(128\sqrt{\pi} r_{0}^{2} |a_{12}|^{5/2})]^{1/3}$, respectively.

The so derived dimensionless GPE can be written as
\begin{equation}
i\frac{\partial \psi }{\partial t}=\left[ -\frac{1}{2}\nabla ^{2}+R(r)+|\psi
|^{3}\right] \psi .  \label{Eq7}
\end{equation}%
It conserves norm $N$, energy $E$, and the angular momentum (in the case of spherically-symmetric potential), whose $z-$component $M_{z}$ is written below:
\begin{equation}
\begin{aligned} N=&\iiint |\psi|^2\text{d}\textbf{r}, \\
E=&\frac{1}{2}\iiint \left[|\nabla\psi|^2+\frac{4}{5}|\psi|^5+2R|\psi|^2\right]\text{d}\textbf{r}, \\ 
M_z=&i\iiint\left[ \psi^*\left(y \frac{\partial}{\partial x}-x \frac{\partial}{\partial y}\right)\psi\right]\text{d}\textbf{r} 
\end{aligned}  \label{Eq8}
\end{equation}

Stationary solutions for hopfions in the LHY superfluid can be defined, in cylindrical coordinates ($\rho ,\theta ,z$), as
\begin{equation}
\psi (\mathbf{r},t)=\phi (\rho ,z)\exp (im\theta )\exp (-i\mu t),
\label{psi}
\end{equation}%
where $\mu $ is the chemical potential and $m=\iint \mathrm{arctan} [\phi_{i}(x,y,z=0)/\phi _{r}(x,y,z=0)]\text{d}x\text{d}y/2\pi $ is the
vorticity (topological charge) of the vertical vortex line, where $\phi_r$ and $\phi_i$ are real and imaginary parts of the complex function $\phi$. Substituting ansatz (\ref{psi}) and considering the isotropic HO potential, $R(\mathbf{r})=\frac{1}{2}%
\omega ^{2}(\rho ^{2}+z^{2})$, in Eq.~(\ref{Eq7}) yields the stationary equation,
\begin{equation}
\frac{1}{2}\left( \frac{\partial ^{2}}{\partial \rho ^{2}}+\frac{1}{\rho }\frac{\partial }{\partial \rho }-\frac{m^{2}}{\rho ^{2}}\right) \phi -\frac{1}{2}\omega ^{2}(\rho ^{2}+z^{2})\phi +\mu \phi -|\phi |^{3}\phi =0.
\label{Eq9}
\end{equation}
Solutions of Eq. (\ref{Eq9}) for the 3D LHY superfluid can be found by means of the Newton-conjugate-gradient \cite{Book2} or relaxation method \cite{Besse}. Below, we fix $\omega \equiv 0.1$ as a typical relevant value.

In contrast to the conventional vortex, with the single vortex line oriented parallel to the $z-$axis and described by real function $\phi $ in Eq. (\ref{psi}), hopfions are represented by complex $\phi \equiv \phi _{r}+i\phi_{i}$, which has nontrivial phase structure due to the presence of the vortex ring,
\begin{equation*}
\theta _{s}\left( x,z\right) =\mathrm{arctan}\frac{\phi _{i}(x,y=0,z)}{\phi_{r}(x,y=0,z)}.
\end{equation*}
The corresponding integer twist topological charge $s$ is then found as 
\begin{equation*}
s=\frac{1}{2\pi }\iint \theta _{s}\left( x,z\right) \text{d}x\text{d}z.
\end{equation*}

\section{Numerical results}

\label{Sec3}
\begin{figure*}[tb]
\centering
\includegraphics[width=0.8\textwidth]{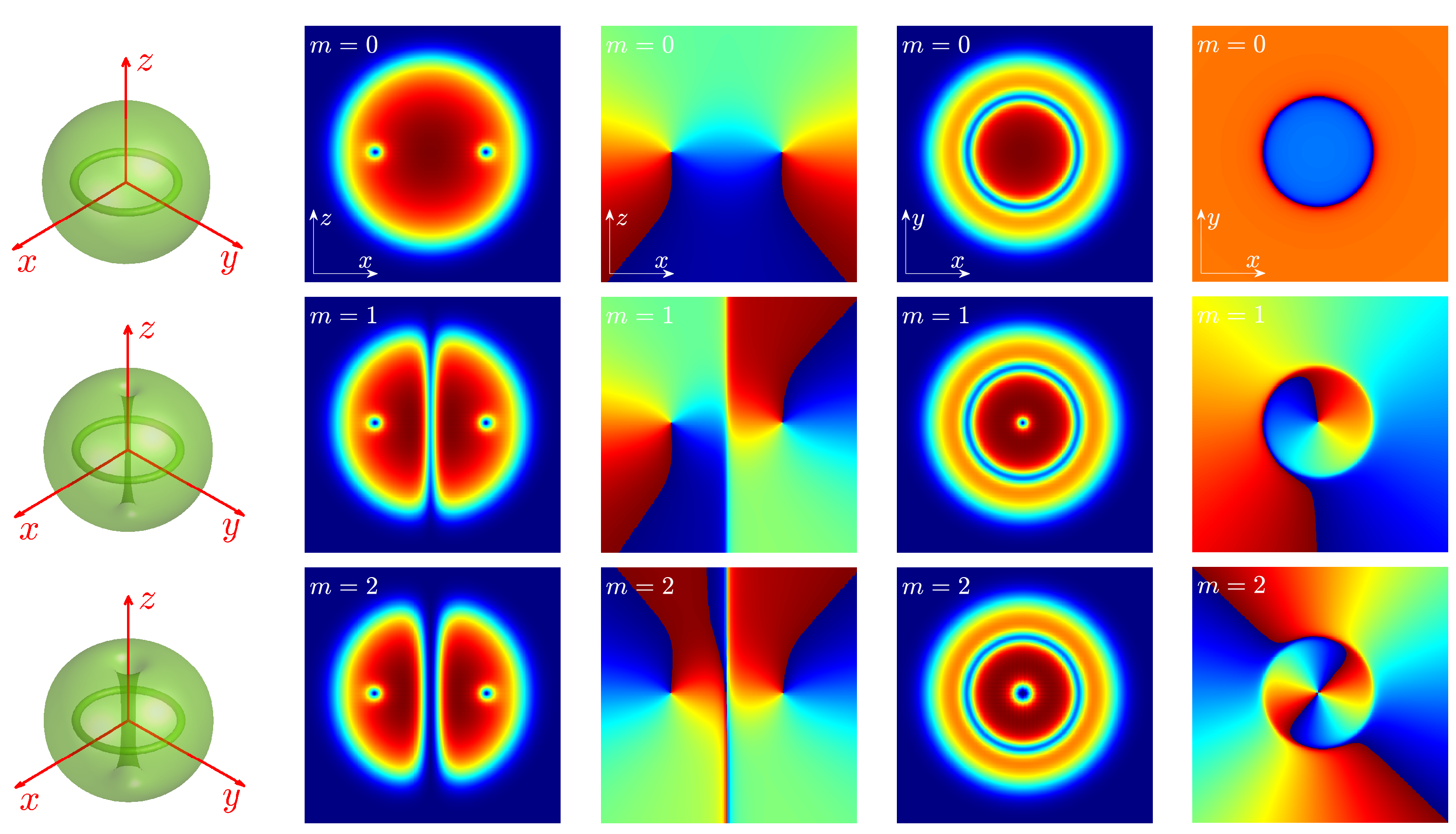}\vskip-0pc
\caption{(a) Top, central, and bottom rows: The hopfions with chemical potential $\protect\mu =1.2$, intrinsic winding number $s=1$, and
vorticities $m=0,1,2$, maintained by the HO potential with $\protect\omega =0.1$. The first column shows the isosurface of $|\protect\phi (x,y,z)|=0.45|\protect\phi _{\text{max}}|$. The second and third columns display $|\protect\phi |$ and $\arg (\protect\phi )$ in the cross section $y=0$. The fourth and fifth columns show $|\protect\phi |$ and $\arg (\protect\phi )$ in the cross section $z=0$. The spatial domain in all the 2D plots is [$-20,+20$].
Here and in other 2D figures, blue and red regions in plots of $|\protect\phi |$ correspond to lower and higher values, respectively.}
\label{fig1}
\end{figure*}

Typical stationary solutions for twisted toroidal states with topological charges $s=1$ and $m=0,1,2$ are presented in Fig.~\ref{fig1}. To the best of
our knowledge, these states are the first reported examples of hopfions in the LHY superfluid. The vortex rings nested in nearly spherically-shaped
hopfions are evident in the figures. The 3D plots demonstrate that the hopfion is based on the toroidal twisted vortex ring coiled around the $z-$axis, which is embedded in the 3D fundamental ($m=0$) or vortex ($m\geq 1$) mode. At $s=0$, the hopfion reduces to the conventional fundamental or conventional vortex mode \cite{DONG2023113728}. The topological phase structure of the twisted vortex ring is displayed in the third column of Fig. \ref{fig1}. At a fixed chemical potential $\mu $, the dark hole surrounding the $z-$axis expands with the increase of $m$. The right column shows clearly the topology of vortex line and the pronounced phase discontinuity between the regions of $\rho >R$ and $\rho <R$, where $R$ is the radius of vortex ring. The $z-$component of the angular momentum of these states is related to the norm as in conventional vortices, \textit{viz}., $M_{z}=mN$. No hopfion modes with $s>1$ have been found, even for $m=0$.

\begin{figure}[tbph]
\centering
\includegraphics[width=0.45\textwidth]{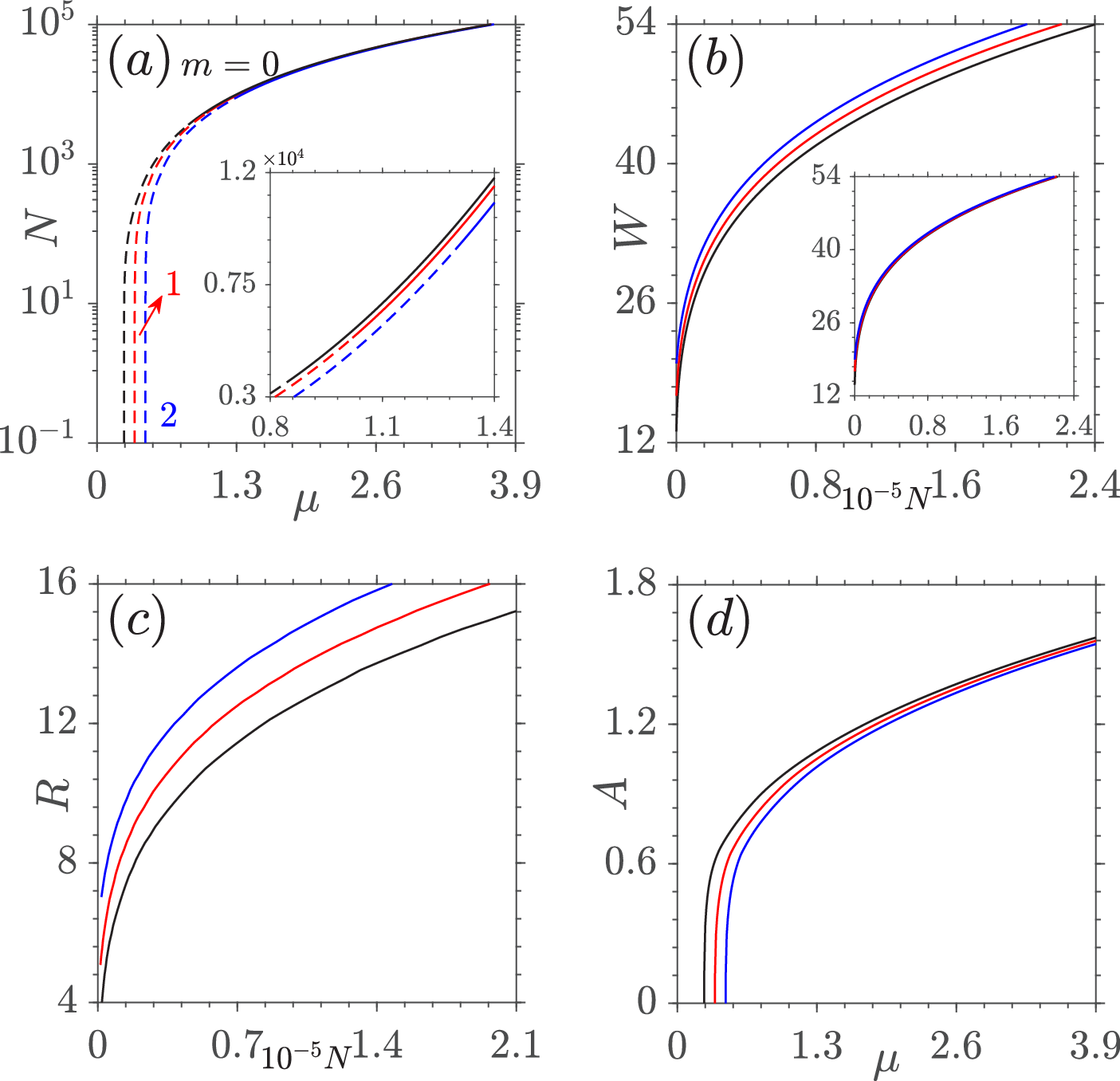}\vskip-0pc
\caption{Families of the twisted toroidal states (hopfions) with intrinsic winding number $s=1$ and different values of vorticity $m$: (a) the norm vs.
the chemical potential; (b) the effective width vs. the norm; (c) radius $R$ of the coiled vortex ring vs. the norm. (d) The dependence of the hopfion's
amplitude, $A=\text{max}(|\protect\phi |)$, on $\protect\mu $. The inset in (a) is the zoom-in, taken close to the border points separating stable
(solid) and unstable (dashed) subfamilies. For the clarity's sake, the curves with $m=0$ and $2$ (black and blues ones, respectively) in (b, c) are
shifted by $\Delta W(\Delta R)=-1$ and $+1$, respectively. }
\label{fig2}
\end{figure}

Families of the hopfions with $s=1$ and $m=0,1,2$ are represented, in Fig.~\ref{fig2}(a), by the dependence of the norm $N$ on the chemical potential $\mu $, which shows monotonous increase of $N$ with the growth of $\mu $. At fixed $\mu $, the hopfions with smaller values of vorticity $m$ possess
(slightly) higher norms. The hopfions cease to exist below a cutoff value $\mu _{\text{cut}}$ of the chemical potential, where $N$ vanishes, and they transform into linear modes of the spherical HO potential, which makes it possible to find $\mu _{\text{cut}}=(m+s+3/2)\omega $, as
eigenvalues of the 3D HO potential.

The size of the hopfion is characterized by its effective width, defined by
\begin{equation}
W^{2}\equiv \frac{\iiint (x^{2}+y^{2}+z^{2})|\phi|^{2}\text{d}x\text{d}y\text{d}z}{N}.  \label{W}
\end{equation}%
It exhibits a monotonic increase with the increase of $N$, indicating spatial expansion of the hopfions in condensates with larger atomic
populations [Fig.~\ref{fig2}(b)]. The hopfions with different values of $m$ demonstrate nearly identical effective widths\ for fixed $N$, as shown in
the inset of Fig.~\ref{fig2}(b), where the hopfions with higher vorticity $m$ exhibit a negligibly small increase of $W$ in comparison with lower values
of $m$. To enhance visual clarity in the main panel, we vertically offset the curves for $m=0$ and $m=2$ downward and upward by $\Delta W=1$,
respectively.

Figure~\ref{fig2}(c) shows that the radius of the vortex ring increases with norm $N$, in contrast with the hopfions maintained by
inhomogeneous self-repulsive media \cite{PhysRevLett.113.264101}, where the radius of the coiled vortex thread slowly decreases with $N$, remaining
finite (non-collapsing) as $N\rightarrow \infty $. The hopfions' amplitude also displays monotonous dependence on $\mu$ for different values of $m$.
This means that the maximum density of the atomic cloud increases with the growth of $\mu$.

The stability of the hopfions is the critically significant issue, as it determines their experimental observability of the modes. To explore the
stability, we consider perturbed modes, looked for as
\begin{eqnarray}
&&\psi (\rho ,\theta ,z,t)=[\phi (\rho ,z)+f(\rho ,z)\exp (\delta t+ik\theta
)+  \notag  \label{Eq10} \\
&&g^{\ast }(\rho ,z)\exp (\delta ^{\ast }t-ik\theta )]\exp (-i\mu t+im\theta
),  \notag
\end{eqnarray}%
where $f$ and $g$ are the infinitesimal perturbations with the (generally, complex) growth rate $\delta $, the stability condition being that all
eigenvalues must have Re$(\delta )\leq 0$, while $k$ is an integer representing the angular index of the perturbations, and the asterisk stands
for the complex conjugate. The substitution of the perturbed solutions in Eq.~(\ref{Eq7}) leads to the linearized Bogoliubov -- de Gennes equations
\cite{PhysRevA.98.013612}, for the perturbations $f(\rho ,z)$ and $g(\rho ,z)$:
\begin{equation}
\begin{split}
i\delta f=& -\frac{1}{2}\left( \frac{\partial ^{2}}{\partial \rho ^{2}}+\frac{1}{\rho }\frac{\partial }{\partial \rho }+\frac{\partial ^{2}}{\partial z^{2}}-\frac{(k+m)^{2}}{\rho ^{2}}\right) f \\
& +\left( \frac{5}{2}|\phi |^{3}+V-\mu \right) f+\frac{3}{2}|\phi |\phi^{2}g,
\\
i\delta g=& ~~~~\frac{1}{2}\left( \frac{\partial ^{2}}{\partial \rho ^{2}}+\frac{1}{\rho }\frac{\partial }{\partial \rho }+\frac{\partial ^{2}}{\partial z^{2}}-\frac{(k-m)^{2}}{\rho ^{2}}\right) g \\
& -\left( \frac{5}{2}|\phi |^{3}+V-\mu \right) g-\frac{3}{2}|\phi
|(\phi^{\ast })^{2}f.
\end{split}
\label{Eq11}
\end{equation}%
Equations (\ref{Eq11}) can be solved by means of the Fourier collocation method \cite{Book2}.

\begin{figure}[tbph]
\centering
\includegraphics[width=0.48\textwidth]{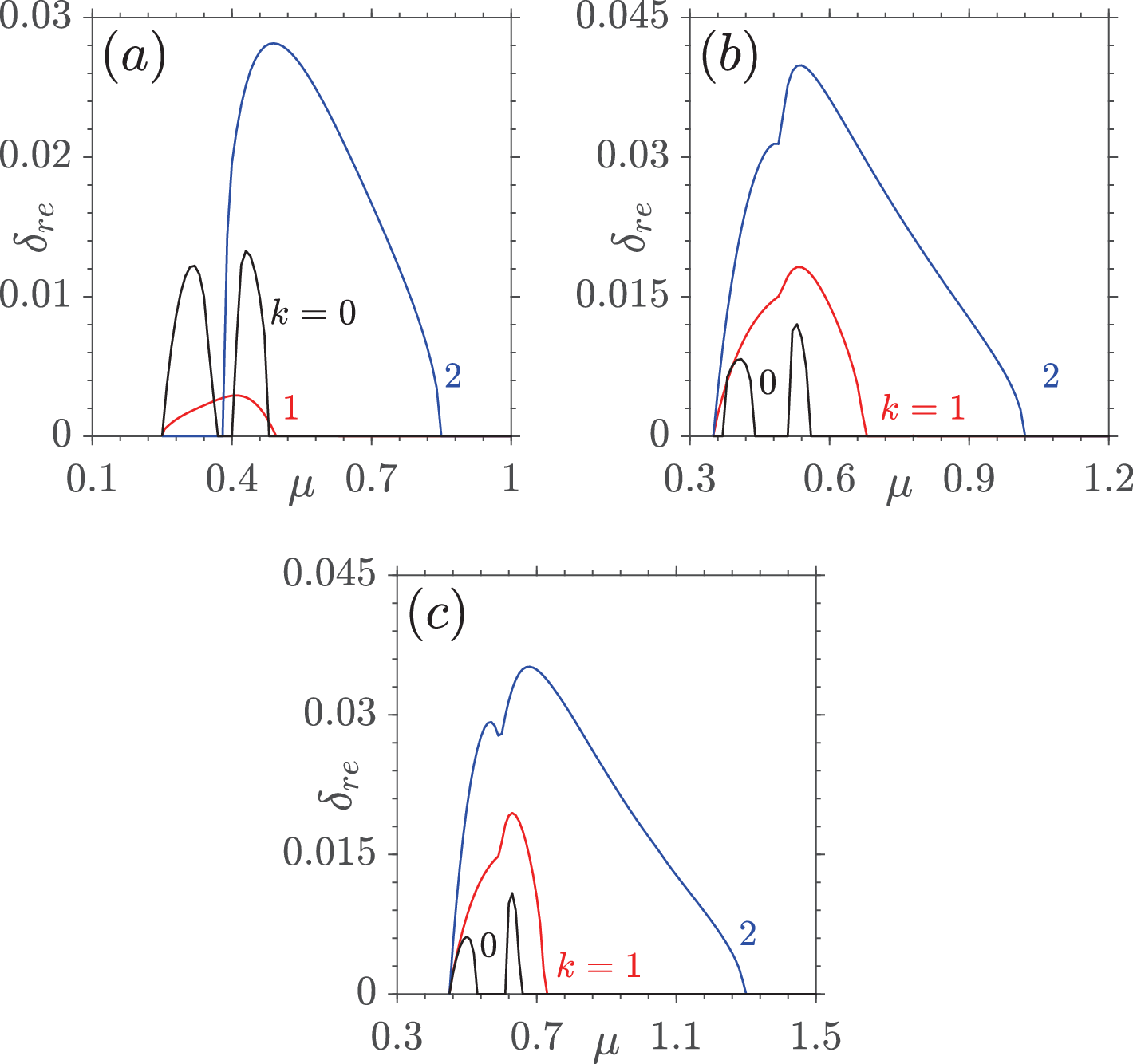}\vskip-0pc
\caption{The real part of the instability growth rate, for different values of azimuthal perturbation index $k$, versus the chemical potential. The
instability domains are $\protect\mu \in [0.25, 0.85]$ for $m=0$ (a), $[0.35, 1.02]$ for $m=1$ (b) and $[0.45, 1.30]$ for $m=2$ (c). }
\label{fig3}
\end{figure}

The numerical solution of Eqs.~(\ref{Eq11}) produces the stability regions for the hopfions with $s=1$ and $m=0,1,2$, according to Fig.~\ref{fig3},
where the instability characteristics exhibit strong dependence of the azimuthal index $k$ of the perturbations. Each value $k=0,1,2$ produces
drastically different dependences of Re$\left( \delta \right) $ on the hopfions's chemical potential $\mu $. The global instability domain is
obtained of all particular ones corresponding to different values of $k$. For the hopfions with $m=1$ and $2$, the instability spectrum is entirely
dominated by the perturbations with $k=2$ [Figs.~\ref{fig3}(b,c)], while the hopfions with $m=0$ show the dominant instabilities with $k=0$ and $1$ at
low norms, in addition to the leading instability with $k=2$ at larger norms [Fig.~\ref{fig3}(a)]. The overall instability intervals are $\mu \in \lbrack
0.25,0.85]$, $[0.35,1.02]$, and $[0.45,1.3]$ for the hopfions with $m=0,1,$ and $2$, respectively. Actually, the instability growth rates Re$(\delta )>0$
are small for unstable modes, which implies slow evolution of unstable hopfions, as shown below in Fig.~\ref{fig5}.

In contrast to\ the fundamental state ($s=0,m=0$), which is completely stable in its existence domain \cite{DONG2023113728}, the hopfions with $s=1$ exhibit instability at low norms even for $m=0$. Nevertheless, the hopfions with different vorticities $m$ achieve stabilization when their norm exceeds a
threshold value, which is indicated by the solid-dashed-segment transitions in Fig.~\ref{fig2}(a). The stability properties of the hopfions in the 3D
LHY superfluid are in sharp contrast to those in the previously studied inhomogeneous repulsive media, where the hopfions with $s=1,m=0$ and $s=1,m>1 $ are, respectively, completely stable and unstable in their entire existence domains \cite{PhysRevLett.113.264101}.

\begin{figure}[h]
\centering
\includegraphics[width=0.45\textwidth]{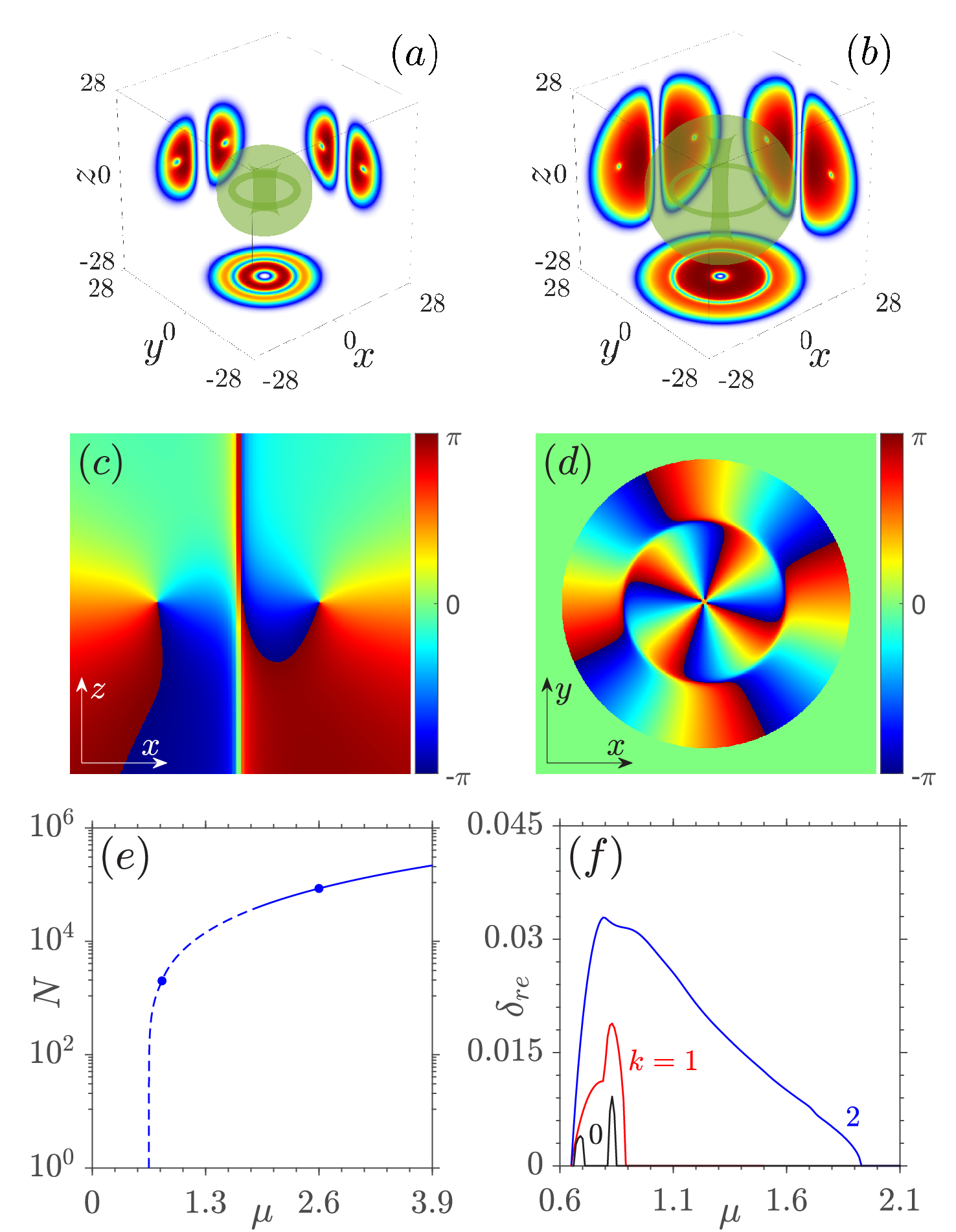}
\caption{Isosurfaces $|\protect\phi \left( x,y,z\right) |=0.55\protect\phi _{\text{max}}$ illustrating profiles of the hopfions with $m=4,s=1$ at $\protect\mu =1.1$ (a) and $2.6$ (b). The 2D profiles on the walls and bottoms of the 3D boxes in (a) and (b) show the wave function in cross sections $x=0$, $y=0$, and $z=0$, respectively. (c, d) Phase distributions of $\protect\phi $ in cross sections $y=0$ and $z=0$. The spatial domains are $(x,z)\in \lbrack -28,+28]$ in (c), and $(x,y)\in \lbrack -28,+28]$ in (d). (e) Norm $N$ vs. chemical potential $\protect\mu $ for the hopfions with $m=4,s=1$. (f) The instability growth rates vs. $\protect\mu $ for different values of azimuthal perturbation index $k$.}
\label{fig4}
\end{figure}

Further, we systematically investigated the hopfions with higher vorticities $m$. Representative examples of such solutions with $s=1,m=4$ are presented in Fig.~\ref{fig4}. The hopfions exhibit characteristic features of twisted toroidal modes [Figs.~\ref{fig4}(a,b)]. The phase structure in the ($x,z$) plane is typical for hopfions, where one can identify the presence of the vortex line and ring [Fig.~\ref{fig4}(c)]. In the ($x,y$) plane, the vortex toroidal ring creates a pronounced phase discontinuity between the regions of $\sqrt{x^{2}+y^{2}}\equiv \rho >R$ and $\rho <R$, with $R$ denoting the radius of the coiled vortex ring [Fig.~\ref{fig4}(d)].

The hopfions with $m=4$ achieve the stability at $N>N_{\text{cr}}\approx 4\times 10^{5}$. The perturbations with azimuthal index $k=2$ still dominate
the instability in this case, the corresponding instability region being $\mu \in \lbrack 0.65,1.93]$. While the theoretical results demonstrate that
the hopfions with $m>4$ may be stabilized at even higher norms, the experimental realization of this prediction may face challenges due to the
difficulty in preparing condensates with so large atom numbers.
\begin{figure}[t]
\centering
\includegraphics[width=0.48\textwidth]{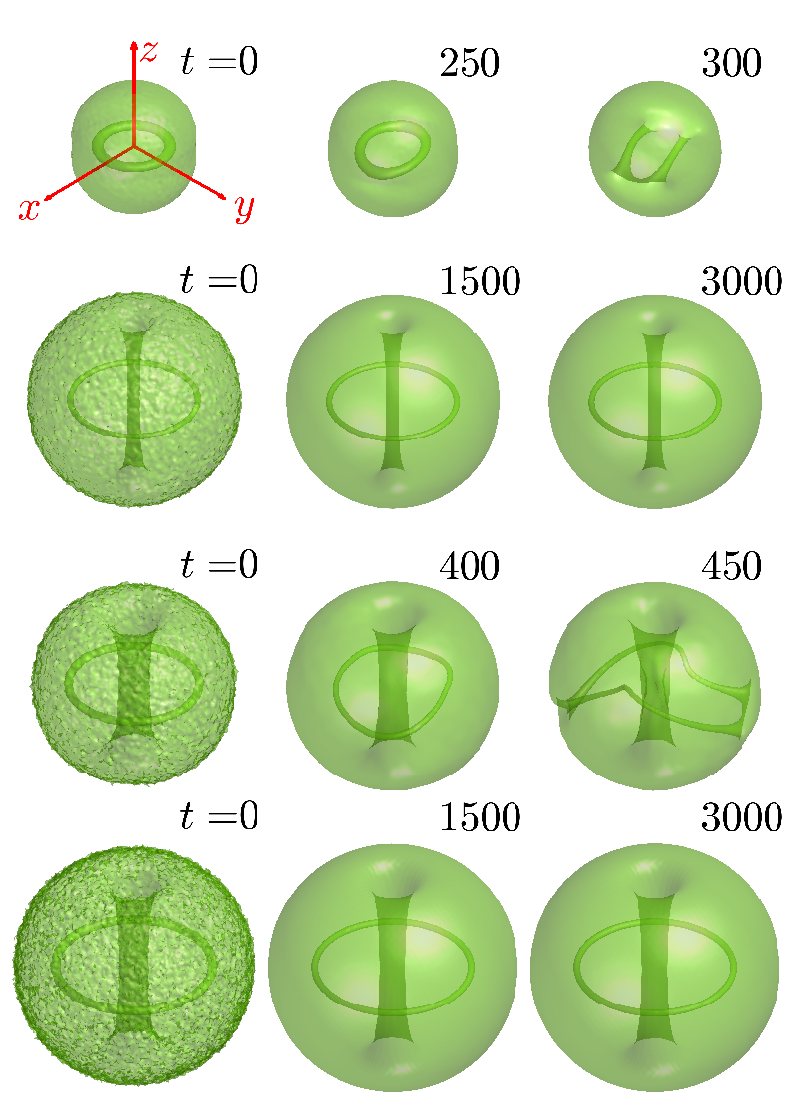}
\caption{Isosurfaces $\left\vert \protect\phi \left( x,y,z\right)
\right\vert =0.35|\protect\phi |_{\text{max}}$ and $0.45|\protect\phi |_{\text{max}}$ in the first three rows, and in the bottom one, show,
respectively, the unstable evolution of the hopfions with $m=0,\protect\mu=0.6$ and $m=4,\protect\mu =1.5$ (the first and third rows, respectively),
and the stable evolution for $m=2,\protect\mu =1.5$ and $m=4,\protect\mu =2.16$ (the second and fourth rows). Broadband random perturbations are
applied to all hopfions at $t=0$. In all cases, the twist topological charge is $s=1$.}
\label{fig5}
\end{figure}

To confirm the above predictions of the linear-stability analysis, we have conducted extensive simulations of the evolution of perturbed hopfions by
dint of the split-step Fourier method. The results for stable hopfions are presented with white noise added to the input at $t=0$. Representative
examples are shown in Fig.~\ref{fig5}. At $\mu =0.6$, the vortex ring of the unstable hopfion with $s=1,m=0$ exhibits obvious deformation at $t=250$, and
is destroyed at $t=300$ (the first row of Fig.~\ref{fig5}). The corresponding maximum instability growth rate is $0.024$ [Fig.~\ref{fig3}(a)] is relatively small, thus allowing nearly stable evolution at $t<200$. We note the instability develops even if noise is not initially added. On the other hand, the stable hopfion with $s=1,m=2$ keeps its vortex line and ring intact in the course of an arbitrary long evolution (the second row of Fig.~\ref{fig5}). The simulation results also show good agreement with the prediction of the linear-stability analysis for the hopfions with higher values of vorticity $m$. For instance, the maximal instability growth rate of the hopfion with $m=4$ at $\mu=1.5$ is $0.0125$, see Fig.~\ref{fig4}(f). This small value expectedly delays the onset of the deformation of the vortex ring. A perturbed stable hopfion with larger norm quickly radiates away the initially added noise and then keeps its robust structure in the course of indefinitely long evolution (see the fourth row of Fig.~\ref{fig5}). It is worthy to note that stable hopfions with $m\geq 2$ in single-component models have not been previously reported.

Finally, we note that the experimental realization of the predicted hopfions can be achieved using the setup proposed in Ref.~\cite{PhysRevLett.126.230404}. In this context, we consider the $^{39}$K atomic gas, which was used in recent experimental configurations \cite{Science2018,PhysRevLett.120.235301,PhysRevLett.126.230404}. Setting $r_{0}=0.3$ $\mathrm{\mu }$m yields a characteristic time scale $t_{0}\sim 55.2$ ms. The scattering lengths are chosen as $|a_{12}|=\sqrt{a_{11}a_{22}}=53a_{0}$ with $a_{22}/a_{11}=2.5$ ($a_{0}$ is the Bohr
radius). The frequency of HO potential is $\omega =\omega _{0}r_{0}\sqrt{m/\varepsilon _{0}}$, where $\omega _{0}$ represents the actual trapping
frequency. The physical atom number $\mathcal{N}$ relates to the dimensionless norm $N$ through $\mathcal{N}=(\varepsilon _{0}r_{0}^{3}/g_{12})N$ with the coupling constant $g_{12}=4\pi \hbar ^{2}a_{12}/m$. Based on these parameters, we estimate feasible values for the proposed experimental realization as $\omega _{0}\sim 288.13\times 2\pi $ Hz, and $\mathcal{N}\sim 4.26\times 10^{4}$, all being within the range of current experimental capabilities.

\section{Conclusion}

\label{Sec4} In this work we have investigated the stationary states and dynamics of the $3$D LHY (Lee-Huang-Yang) superfluid in the regime where the
MF (mean-field) interactions are completely canceled, the LHY nonlinearity being the dominant term. The system is modeled by the effective
single-component GPE (Gross-Pitaevskii equation), featuring zero inter-component interaction and quartic LHY self-repulsion. Hopfions, as 3D
complex topological states, can be created in the corresponding binary BEC if the\ self-repulsion is balanced by the HO (harmonic-oscillator)
potential. These hopfions exhibit coexisting vortex line and vortex ring, characterized by two independent topological
winding numbers: overall vorticity $m$ and twist $s$. We have identified families of stable hopfions with $s=1$ and $m$ ranging from $0$ to $4$. The
norm, effective radius, peak density, and radius of coiled vortex ring all increase with the growth of the hopfion's chemical potential. Although the
instability domain of the hopfions expands with the increase of $m$, such stable states have been found in a wide parameter region (at least, for $m\leq 4$),
provided that the norm exceeds a threshold value.

%省基金 编号
\vskip0.5pc \textbf{CRediT authorship contribution statement} Liangwei Dong:
Conceptualization and writing the original draft; Mingjing Fan: Numerical
calculations; Boris A. Malomed: Analytical investigation and work on the
draft; Yaroslav V. Kartashov: Review, editing and validation.

\vskip0.5pc \textbf{Declaration of competing interest} The authors declare
that they have no known competing financial interests or personal
relationships that could have appeared to influence the work reported in
this paper.

%The authors declare the following financial interests/personal relationships which may be
%considered as potential competing interests: Liangwei Dong reports financial support provided by Natural Science Basic Research Program in Shaanxi
%Province of China (grant No. 2022JZ-02); Boris A. Malomed reports financial support provided by the Israel Science Foundation through grant No. 1695/22.
%Y.V.K. reports financial support provided by the research project FFUU-2024-0003 of the Institute of Spectroscopy of the Russian Academy of Sciences.

\vskip0.5pc \textbf{Data availability} Data will be made available on
request.

\vskip0.5pc \textbf{Acknowledgments} This work is supported by the Natural
Science Basic Research Program of Shaanxi Province of China (Grant No.
2022JZ-02) and Israel Science Foundation (Grant No. 1695/22). Y.V.K.
acknowledges funding by the research project FFUU-2024-0003 of the Institute
of Spectroscopy of the Russian Academy of Sciences.

%\bibliographystyle{plain}

%\bibliography{referenceV4only}
%\bibliographystyle{elsarticle-num-names}
%\bibliographystyle{osajnl}%{unsrt}

%\begin{thebibliography}
%{Junquera et~al.(2023)Junquera, Nahas, Prokhorenko, Bellaiche, Iniguez, Schlom, Chen, Salahuddin, Muller, Martin, and Ramesh}

%\begin{thebibliography}{Junquera et~al.(2023)Junquera, Nahas, Prokhorenko,
%Bellaiche, Iniguez, Schlom, Chen, Salahuddin, Muller, Martin, and Ramesh}
%\bibitem{} \providecommand{\natexlab}[1]{#1} \providecommand{\url}[1]{%
%\texttt{#1}} \providecommand{\urlprefix}{URL } \expandafter\ifx\csname %
%urlstyle\endcsname\relax
%\providecommand{\doi}[1]{doi:\discretionary{}{}{}#1}\else
%\providecommand{\doi}[1]{doi:\discretionary{}{}{}\begingroup
%		\urlstyle{rm}\url{#1}\endgroup}\fi
%\providecommand{\bibinfo}[2]{#2}

\end{document}